\documentclass{imsart}

\usepackage{amssymb, amsthm, amsmath}
\RequirePackage{natbib}

\usepackage{bm}							
\usepackage[ruled,vlined]{algorithm2e} 	
\usepackage{graphicx} 					
\usepackage{color} 						
\usepackage{colortbl}					
\usepackage{float}						
\usepackage{fancyhdr}
\usepackage{enumitem}
\usepackage{soul} 

\definecolor{Grey}{gray}{0.9}			

\usepackage{draftwatermark}
\SetWatermarkText{\textcopyright \ 2019 SDSU}
\SetWatermarkScale{3}

\startlocaldefs
\endlocaldefs

\begin{document}

\begin{frontmatter}

\title{Quantification of the weight of fingerprint evidence using a ROC-based Approximate Bayesian Computation algorithm for model selection \thanksref{t1}}
\thankstext{t1}{This project was supported in part by Award No. 2014-IJ-CX-K088 awarded by the National Institute of Justice, Office of Justice Programs, U.S. Department of Justice, and in part by the National Science Foundation under Grant DMS-1127914 to the Statistical and Applied Mathematical Sciences Institute. The opinions, findings, and conclusions or recommendations expressed in this paper are those of the authors and do not necessarily reflect those of the Department of Justice or the National Science Foundation. We would like to thank Dr. Danica Ommen for her help in the initial implementation of ABC for forensic evidence.}
\runtitle{ROC-based ABC algorithm for fingerprint evidence}


\author{\fnms{Jessie} \snm{Hendricks}\ead[label=e1]{jessiehhendricks@gmail.com}}
\and
\author{\fnms{Cedric} \snm{Neumann}\ead[label=e2]{cedric.neumann@sdstate.edu}}
\and
\author{\fnms{Christopher P.} \snm{Saunders}\ead[label=e3]{christopher.saunders@sdstate.edu}}

\address{Department of Mathematics and Statistics\\ South Dakota State University\\ Brookings, SD, USA\\\printead{e1,e2,e3}}

\runauthor{Hendricks et al.}

\begin{abstract}
For more than a century, fingerprints have been used with considerable success to identify criminals or verify the identity of individuals. The categorical conclusion scheme used by fingerprint examiners, and more generally the inference process followed by forensic scientists, have been heavily criticised in the scientific and legal literature. Instead, scholars have proposed to characterise the weight of forensic evidence using the Bayes factor as the key element of the inference process. In forensic science, quantifying the magnitude of support is equally as important as determining which model is supported. Unfortunately, the complexity of fingerprint patterns render likelihood-based inference impossible. In this paper, we use an Approximate Bayesian Computation model selection algorithm to quantify the weight of fingerprint evidence. We supplement the ABC algorithm using a Receiver Operating Characteristic curve to mitigate the effect of the curse of dimensionality. Our modified algorithm is computationally efficient and makes it easier to monitor convergence as the number of simulations increase. We use our method to quantify the weight of fingerprint evidence in forensic science, but we note that it can be applied to any other forensic pattern evidence.
\end{abstract}




\tableofcontents

\end{frontmatter}

\section{Introduction}
\label{BF.fingerprint.evidence}

For more than a century, fingerprints have been used with considerable success to identify criminals or verify the identity of individuals. In this paper, we define a \textit{fingermark} (left panel of Figure \ref{fig:fingerprint}), or \textit{mark}, as the impression resulting from the inadvertent contact between the finger of an unknown donor and a surface (e.g., at a crime scene). We define a \textit{control print} (right panel of Figure \ref{fig:fingerprint}), or \textit{print}, as a finger impression collected under controlled conditions from a known donor (e.g., a suspect). The purpose of forensic fingerprint examination is to support the inference of the donor of a fingermark. Currently, this inference process relies on the visual comparison between the fingermark and control prints from one or more candidate donors who may have been selected through a police investigation or by searching the fingermark in a database of prints from known individuals. 

\begin{figure}[H]
	\centering
	\includegraphics[width=0.8\textwidth]{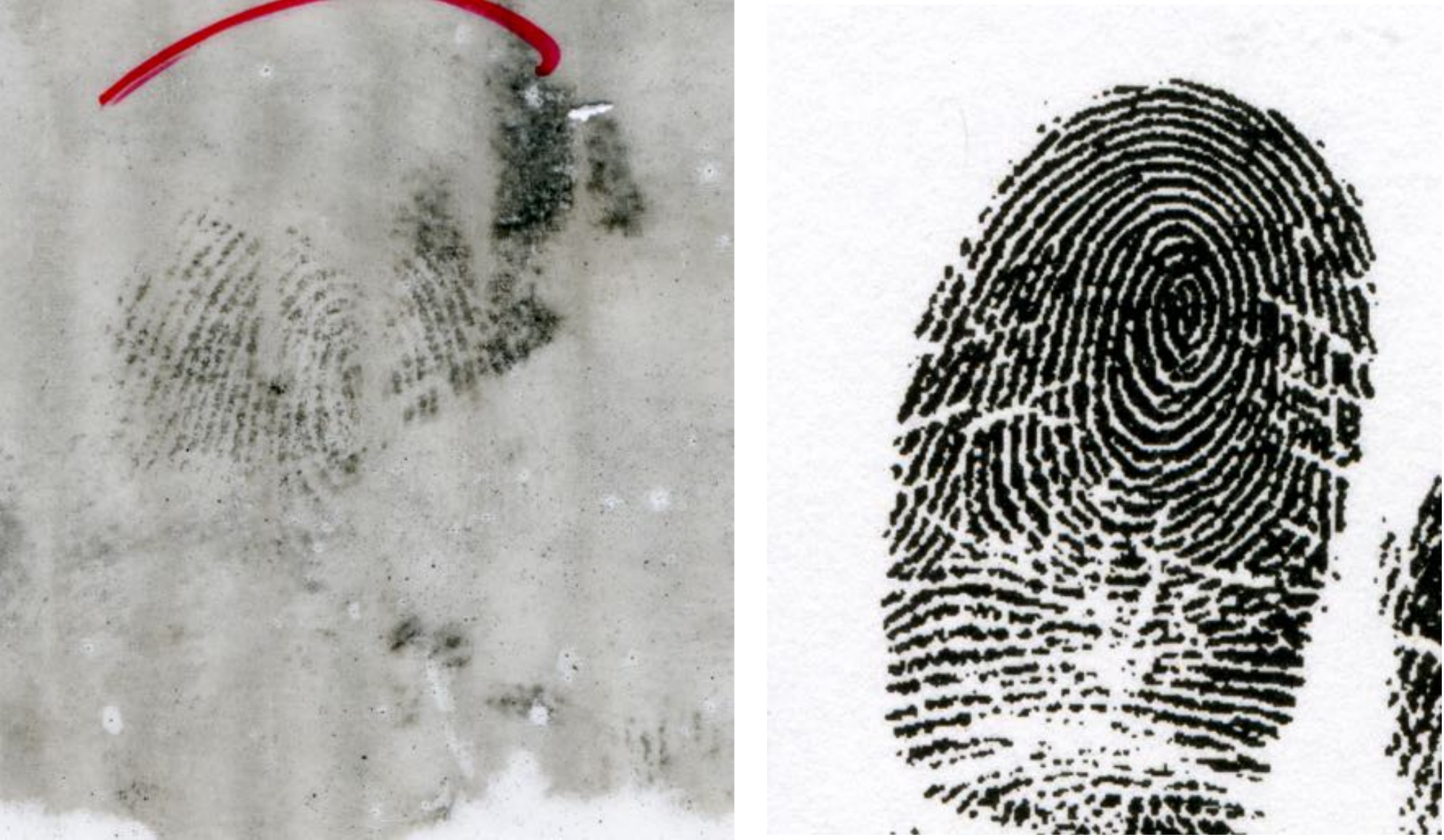}
	\caption{\label{fig:fingerprint} Left panel: latent print. Right panel: control print. Ridges appear darker than background. Both impressions were made by the same finger. Their comparison shows that both ridge flows are affected by different distortion and degradation effects.}
\end{figure}

Following the examination of fingerprint evidence, it is customary for the examiner to report one of two possible conclusions: an opinion of ``identification", implying that the source of the fingermark is the donor of a given control print; or an opinion of ``exclusion", implying that the source of the fingermark is not a considered candidate. Alternatively, the examiner may find the examination ``inconclusive", indicating that the characteristics of the impressions being compared are not sufficient to reach one of the two possible conclusions (e.g., when the impressions are too small or too degraded).

The categorical conclusion scheme used by fingerprint examiners, and more generally the inference process followed by forensic scientists, have been heavily criticised in the scientific and legal literature \citep{Cole:2004, Cole:2005, Cole:2009, Kaye2003, Saks2005, SaksKoehler:2008, Zabell:2005}. Instead, scholars have proposed to characterise the weight of forensic evidence using the Bayes factor as the key element of the inference process (see \citet{Aitken2010} for a comprehensive discussion).

It is worth stressing that forensic scientists do not have a complete picture of all the evidence available in a given criminal case (e.g., eyewitness evidence; means, motive and opportunity of an individual of interest; other material and non-material evidence), which prevents them from assigning probabilities to the different propositions that the parties may have regarding a particular criminal activity. Furthermore, forensic scientists are not tasked with the fact-finding mission in the criminal justice system and are not in charge of the decision-making with respect to the propositions of the parties. Therefore, the role of the scientist is necessarily limited to reporting the amount of support of the forensic evidence for these propositions in the form of a Bayes factor. It is of primary importance to note that, in this setting, we are not only concerned with supporting the correct model, but also in supporting it with the appropriate magnitude. An appropriate magnitude of support is critical to ensure the coherent combination of the respective weight of the multiple pieces of evidence that may be considered in a given case. This imperative requirement is the main motivation behind the present work and is the main force that drives us away from more deterministic pattern recognition algorithms.

Assuming a fingermark recovered in connection with a crime and a suspect, Mr. X., we consider the following two alternative propositions:
\begin{itemize}
	\item [$H_1:$] the fingermark was left by Mr. X.
	\item [$H_2:$] the fingermark was not left by Mr. X., but by another person in a relevant population of potential donors.
\end{itemize}
To address the so-called prosecution and defence propositions, $H_1$ and $H_2$, we define two corresponding models: $M_1$, representing how Mr. X. generates fingermarks, and $M_2$, representing how fingermarks are generated by the donors in the population of alternative sources. Let $\bm{e}_u$ denote the set of observations made on the fingermark. We are interested in evaluating the following Bayes factor:
\begin{equation}
\label{eq:eq1}
	BF = \dfrac{\pi(M_1|\bm{e}_u)}{\pi(M_2|\bm{e}_u)} \cdot \dfrac{\pi(M_2)}{\pi(M_1)} = \dfrac{f(\bm{e}_u | M_1)}{f(\bm{e}_u| M_2)},
\end{equation}
where $\pi(\cdot)$ is a measure of belief about the model and its parameters (see \citet{Robert2007}, chapter 7, for a formal discussion on Bayesian model selection), and $f(\cdot)$ represents a (marginal) probability density function.

Fingerprints are usually characterised through certain features of the ridge pattern, such as the general pattern formed by the friction ridge flow and events disturbing the continuity of the ridges. These events, traditionally called minutiae, occur when a ridge ends (ridge ending) or bifurcates (bifurcation). Other types of events exist but are mainly combinations of these two basic types. 

\begin{figure}[H]
	\centering
	\includegraphics[width=0.5\textwidth]{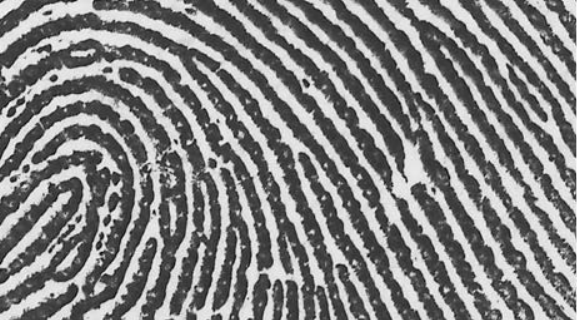}
	\caption{\label{fig:fingerprint} Details of the ridge pattern. Ridges appear darker than background. Ridges can end (\textit{ridge ending}), or split (\textit{bifurcations}); ridges can create enclosures (in upper left corner), or can be very short (upper right corner); ridges can be so short that they appear as \textit{dots}. Short, narrow and non-continuous ridges that appear between two parallel ridges are called \textit{incipient ridges}. White dots within the ridges are sweat pores.}
\end{figure} 

When comparing two fingerprints, an examiner first verifies the compatibility of their general patterns and then determines whether the spatial relationships, types, and orientations of the minutiae on both impressions correspond. Mathematically characterising fingerprint patterns results in heterogeneous, high-dimension random vectors: minutia locations and spatial relationships are continuous measurements; minutia types are discrete observations; and minutia orientations are circular measurements. In addition, the dimensionality of the problem increases with the number of minutiae observed on a given impression. Therefore, modelling the joint likelihood of the characteristics of multiple features observed on an impression seems to be an unreasonable challenge.

While many attempts have been made to assign Bayes factors for fingerprint evidence (see \citet{Abraham2013} for a recent review), no algorithm has gained wide acceptance. In particular, while the results obtained with the algorithm proposed by \citet{Neumann2012} were used to support the general admissibility of fingerprint evidence in U.S. courts (State v. Hull, \citeyear{State.v.Hull}; State v. Dixon, \citeyear{State.v.Dixon}), commentators argued that the model had two main shortcomings. Some commentators discussed that the algorithm did not result in a formal Bayes factor as it does not formally incorporate user beliefs \citep{Cowell2012, Graversen2012, Kadane2012, Lauritzen2012, Stern2012}. Others noted that the algorithm relied on an ad-hoc weighting function used to palliate the inability of the authors to assign Bayes factors \textit{at} 0, and that this function had no other justification than convenience \citep{Balding2012, Fotopoulos2012, Jandhyala2012, Kadane2012}. 

In this paper, we take advantage of the similarities between the algorithm proposed by \citet{Neumann2012} and the Approximate Bayesian Computation framework to provide a method to formally and rigorously assign Bayes factors to fingerprint evidence. Our method leverages the property of the well-known Receiver Operating Characteristic (ROC) curve to address shortcomings in existing ABC algorithms. Our application addresses the issues raised in relation to \citet{Neumann2012} and provide a much needed general framework for the quantification of the weight of any type of forensic pattern evidence, as long as a similarity measure can be defined to compare two pieces of evidence. 

\section{Approximate Bayesian Computation}

Approximate Bayesian Computation (ABC) originated as a class of algorithms designed to sample from the approximate posterior density of a vector of parameters, $\bm{\theta}$, given an observed data set, $\bm{D}$, without direct evaluation of the likelihood function, $f(\bm{D}|\bm{\theta})$. This class of algorithms is especially useful in complex and high-dimensional settings where the likelihood function is not available in a usable form (see \cite{Robert2011} or \cite{Sisson2019} for an overview).

To sample from an approximate posterior density, vectors of parameter values are first sampled from a prior distribution over the parameter space, and then used to generate pseudo-observations (each denoted $\bm{D}^{\ast}$) from an assumed generative model. A vector of parameter values is retained if the distance measured by a kernel function, $\Delta\{\cdot,\cdot\}$, between values of a summary statistic, $\eta(\cdot)$, of $\bm{D}$ and $\bm{D}^{\ast}$ is less than some tolerance, $t>0$. In other words, the $i^{th}$ sampled parameter vector, $\theta^{(i)}$, is retained if the corresponding distance satisfies $\Delta\{\eta(\bm{D}),\eta(\bm{D}^{\ast(i)})\}<t$.

In some application problems, such as forensic science, statisticians are not concerned with assigning posterior probability distributions in the parameter space, but are interested in performing model selection. However, performing likelihood-based inference using patterns of forensic interest, such as fingerprints, shoe sole impressions, or striations on bullets, is not feasible in the original feature space of the data. Starting with \cite{Pritchard1999}, ABC has evolved into an algorithm that can be used for model selection by considering a model index parameter, $\mathcal{M}$, and its prior distribution over model indices. The model index determines the prior distribution over the parameter space and the likelihood structure used to generate pseudo-observations. The $i^{th}$ sampled model index parameter, $\mathcal{M}^{(i)}$, is retained if the corresponding score satisfies $\Delta\{\eta(\bm{D}),\eta(\bm{D}^{\ast(i)})\}<t$. 

The original ABC method for assigning a posterior probability to a model, given the observed data, is a function of the ratio of the number of times that model index has been retained over the total number of times any model was retained \citep{Pritchard1999, Beaumont2008, Grelaud2009, ToniStumpf2010, Didelot2011, Robert2011}. Mathematically, the ABC posterior odds for the comparison between two models is given by:
\begin{equation}
\label{eq:eq2}
	\dfrac{\pi_{t}(\mathcal{M}=1|\bm{D})}{\pi_{t}(\mathcal{M}=2|\bm{D})}
	 = \lim_{N \to \infty} \dfrac{\sum^N_{i=1}\mathbb{I}(\mathcal{M}^{(i)}=1)\cdot\mathbb{I}(\Delta\{\eta(\bm{D}),\eta(\bm{D}^{\ast(i)})\}\leq t)}{\sum^N_{i=1}\mathbb{I}(\mathcal{M}^{(i)}=2)\cdot\mathbb{I}(\Delta\{\eta(\bm{D}),\eta(\bm{D}^{\ast(i)})\}\leq t)},\\
\end{equation} 
where $\mathbb{I}(\cdot)$ is the indicator function and the subscript $t$ in $\pi_t$ indicates that this measure is a function of the choice of the tolerance level. 

The ABC Bayes Factor, $BF_{abc}$, can then be assigned using the ABC approximation of the posterior odds, divided by the prior odds   
\begin{equation}
\label{eq:eq3}
	BF_{abc}=\frac{\pi_t(\mathcal{M}=1|\bm{D})}{\pi_t(\mathcal{M}=2|\bm{D})}\cdot\frac{\pi(\mathcal{M}=2)}{\pi(\mathcal{M}=1)}.
\end{equation}

The ABC algorithm for comparing two models is summarised in Algorithm \ref{alg:Modified} \citep{Robert2011}. 
\begin{algorithm}
 \KwData{Observed data, $\bm{D}$.}
 \KwResult{ABC Bayes factor, $BF_{abc}$.} 
 \For{i = 1 to $N$}{
 	Sample a model index $\mathcal{M}^{(i)}$ from the model prior, $\pi(\mathcal{M}=m)$, where $m=1,2$;
 	Sample a vector of parameters, $\bm{\theta}^{(i)}$, from the prior density, $\pi(\bm{\theta}| \mathcal{M}^{(i)})$\;
 	Generate a pseudo-observation, $\bm{D}^{\ast(i)}$, from the assumed likelihood, $f(\bm{D}^{\ast}|\bm{\theta}^{(i)})$\;
 	Compute $\Delta\{\eta(\bm{D}),\eta(\bm{D}^{\ast(i)})\}$\;
 }
 Calculate the ratio of counts as defined in Equation (\ref{eq:eq2})\;
 Assign $BF_{abc}$ as in Equation (\ref{eq:eq3})\;
 \caption{ABC model selection algorithm \citep{MarinRobert2014}} 
 \label{alg:Modified}
\end{algorithm}

The benefits of Approximate Bayesian Computation methods are not without cost. Model selection using ABC is subject to two primary sources of error: 
\begin{enumerate}
	\item the quality of the approximation to the likelihood function due to the use of the tolerance, $t$, 
	\item and the loss of information engendered by using a non-sufficient summary statistic.
\end{enumerate}

The effect of the tolerance level, $t$, on the performances of ABC algorithms has been widely discussed since the inception of ABC methods. In short, if $t$ is too large, too many samples from the prior distribution are accepted and the approximation becomes invalid; and, if $t$ is too small the rate of acceptance is too small to produce a stable posterior distribution \citep{Didelot2011, Robert2011}. 

The motivation for using summary statistics, rather than the full data set, stems from the curse of dimensionality encountered with the high-dimensional data sets that are common to scenarios in which ABC is necessary. Ideally a summary statistic that is sufficient across models should be used to enable the convergence of $BF_{abc}$ to the Bayes factor \citep{Robert2011, Didelot2011}. In general, it seems difficult, if not impossible, to find sufficient summary statistics for the type of data that are typically used in ABC method. The goal is then to avoid the curse of dimensionality while minimising the loss of information encountered when using non-sufficient statistics. While it may seem that a general solution addressing the issue of the potential lack of sufficiency of summary statistics could consist in using a large set of summary statistics with the hope that they will jointly tend towards sufficiency by decreasing the loss of information, this is not the case. As the dimensionality of the set of summary statistics increases, the algorithm will also suffer from the curse of dimensionality \citep{Blum2010}. 


Several modifications of the traditional ABC algorithm for model selection have been proposed with the goal of addressing some of the issues related to the determination of the tolerance and the selection of low dimension summary statistics that minimise the loss of information. Most of these modified algorithms are rooted in the one proposed by\citet{Beaumont2008}, who suggests using a polychotomous weighted logistic regression model that has been trained on the summary statistics of the pseudo-observations and corresponding model indices to predict the posterior probability of a model. Modifications include variable selection \citep{Estoup2012}, replacement of the logisitic regression by artificial neural networks \citep{Blum2010} or random forest algorithm \citep{Pudlo2015}, or the use of posterior probabilities as summary statistics \citep{Prangle2014a}.

Finally, we highlight the work by \citet{Marin2014}, who propose a set of necessary and sufficient conditions for summary statistics that result in a corresponding $BF_{abc}$ that supports the true model asymptotically. \citet{Marin2014} suggest that summary statistics should be chosen so their asymptotic mean is different under opposing models - so different that their ranges are non-overlapping. We note that showing that summary statistics satisfy these conditions in a real application is not trivial, and furthermore, that we are not only interested in supporting the correct model, but we are interested in supporting it with the correct magnitude. Nevertheless, we will show below that our method relies on a similar argument to that of \citet{Marin2014}. 

The literature shows that the issues of sufficiency, variable selection and tolerance level are usually entangled. The solutions to these issues proposed to date, and summarised above, have their own sets of complications:
\begin{enumerate}
	\item None of these solutions enables formal monitoring of the convergence of $BF_{abc}$ to the Bayes factor as the value of the tolerance, $t$, is reduced, or as the number of samples increases. Some convergence results for the regression adjustments in the context of parameter inference can be found in \cite{Blum2010} but are not directly applicable to the model selection context. 
	\item Most of these solutions are focusing on selecting the correct model, but are not necessarily designed to support the selected model with the appropriate magnitude, which is a critical requirement in forensic science. 
	\item Most of these methods require to generate summary statistics from pseudo-observations that are as close as possible to the summary statistics of the observed data. That is, these methods attempt to estimate the behaviour of the model selection algorithm as $t \rightarrow 0^+$ without ever observing $t=0$ (or any value even reasonably close due to the curse of dimensionality). This involves the use of potentially complicated variable selection methods, and the definition of an appropriate kernel function to compare summary statistics~\citep{Prangle2017}.  
	\item Methods based on generalised linear models \citep{Beaumont2008}, artificial neural networks \citep{Blum2010}, random forests \citep{Pudlo2015} or other classifiers can be very computationally intensive depending on the dimension of the summary statistics or the number of pseudo-observations generated from the considered models. They also heavily rely on appropriately estimating the (very large number of) parameters of these classifiers. 
	\item Some of the model selection algorithms, such as the one proposed by \citet{Beaumont2008}, are trained using a limited subset of pseudo-observations. These pseudo-observations are selected or weighted based on the proximity of their summary statistics with those of the observed data. Unfortunately, this results in replacing user-defined prior probabilities on model indices with probabilities assigned in unpredictable ways by the algorithm based on the proportions of training data selected from each model. 
\end{enumerate}

In this paper, we propose a modification to the ABC algorithm that utilises a relationship between the ABC Bayes factor and the Receiver Operating Characteristic (ROC) curve. Our novel ROC-based approach addresses the instability created by the use of a heuristic tolerance level. Our algorithm can accommodate sets of summary statistics of any size without being subject to the curse of dimensionality or worrying about the proximity of the summary statistics of the observed and pseudo-data, as long as conditions similar to the ones suggested by \citet{Marin2014} are satisfied.  Critically, our solution enables us to rigorously control its convergence as the number of simulation increases. Furthermore, for the two proposed versions of our algorithm, one requires estimation of only four parameters, and the other does not require estimation of any parameters.



\section{ROC-ABC algorithm for model selection}
\label{ROCmodifiedABC}

ROC curves are traditionally used to measure the performance of a binary classifier as the decision threshold, $t$, is varied across the domain of (dis)similarity scores, $\Delta\{\cdot, \cdot\}$, that can be returned by the classifier \citep{Pepe2003}. ROC curves are obtained by plotting the rate of correct decisions in favour of the first model against the rate of incorrect decisions (false positives) in favour of the first model (i.e. when the second model is true) for all possible values of the decision threshold. 

Defining $F(\cdot)$ and $G(\cdot)$ as the cumulative distribution functions of scores under models 1 and 2, respectively, the general form of the ROC is given by \citep{Pepe2003} as $\text{ROC}(p)~=~F(G^{-1}(p))$,
%
%
where $p$ denotes the rate of false positives in favour of the first model and $G^{-1}(\cdot)$ denotes the quantile function \citep{vanderVaart1998} for scores under the second model. Assuming equal priors for the model indices, we can show that the Bayes factor, $BF_{\eta}$, in Equation~(\ref{eq:eq4}), is a function of the ROC curve constructed on the set of $\Delta\{\eta(\bm{D}), \cdot \}$ from model 1 and the set of $\Delta\{\eta(\bm{D}), \cdot \}$ from model 2:
\begin{align}
	\label{eq:eq7}
	BF_{\eta} & = \lim_{\substack{ t \to 0^+}}BF_{abc}\\
	\label{eq:eq7.5}
	& = \lim_{\substack{ t \to 0^+ \\ N \to \infty}}\dfrac{\sum^N_{i=1}\mathbb{I}[\mathcal{M}^{(i)}=1]\cdot\mathbb{I}[\Delta\{\eta(\bm{D}),\eta(\bm{D}^{\ast(i)})\}\leq t]}{\sum^N_{i=1}\mathbb{I}[\mathcal{M}^{(i)}=2]\cdot\mathbb{I}[\Delta\{\eta(\bm{D}),\eta(\bm{D}^{\ast(i)})\}\leq t]}\cdot \dfrac{\pi(\mathcal{M}=2)}{\pi(\mathcal{M}=1)}\\
	\label{eq:eq8}
	& = \lim_{\substack{ t \to 0^+ \\ N=K+L \to \infty}}\dfrac{\sum^K_{k=1}\mathbb{I}[\Delta\{\eta(\bm{D}),\eta(\bm{D}^{\ast(k)})\}\leq t | \mathcal{M}=1]}{\sum^L_{l=1}\mathbb{I}[\Delta\{\eta(\bm{D}),\eta(\bm{D}^{\ast(l)})\}\leq t|\mathcal{M}=2]}\cdot \dfrac{\pi(\mathcal{M}=2)}{\pi(\mathcal{M}=1)}\\
	\label{eq:eq9}
	& = \lim_{\substack{ t \to 0^+ \\ K \to \infty \\ L \to \infty}}\dfrac{K \cdot \hat{F}_K(t)}{L \cdot\hat{G}_L(t)}\cdot \dfrac{\pi(\mathcal{M}=2)}{\pi(\mathcal{M}=1)}\\
	\label{eq:eq10}
 	& \approx \lim_{t \to 0^+} \dfrac{{F}(t)}{{G}(t)}
	= \lim_{t \to 0^+}\dfrac{{F}({G}^{-1}({G}(t)))}{{G}(t)}\\
	\label{eq:eq13}
	& = \lim_{p \to 0^+}\dfrac{{F}({G}^{-1}(p))}{p}
	= \lim_{p \to 0^+}\dfrac{\text{ROC}(p)}{p}.
\end{align}

Equalities (\ref{eq:eq7}) and (\ref{eq:eq8}) come from \citet{Robert2011}; the equality between (\ref{eq:eq8}) and (\ref{eq:eq9}) is a result of the definition of the empirical distribution function \citep{Wasserman2013}; and, the approximate equality between (\ref{eq:eq9}) and (\ref{eq:eq10}) results from the convergence of the empirical distribution function to the true distribution function as $K$ and $L$ get arbitrarily large, and from the assumption that $K$ and $L$ grow at the same rate given that $\pi(\mathcal{M}=2) = \pi(\mathcal{M}=1)$. 

The relationship expressed between Equations (\ref{eq:eq7}) and (\ref{eq:eq13}) shows that the ABC Bayes factor for two alternative models of interest can be assigned using the ratio between ROC($p$) and $p$ as the rate of false positives in favour of model 1 approaches 0. This notable result allows us to express the convergence of the ABC Bayes factor as a function of the rate of false positives in favour of model 1. Our result has significant practical implications when it comes to using ABC for model selection:
\begin{enumerate}
\item Our solution allows to monitor the convergence of the output of the algorithm to $BF_{\eta}$ as function of a single, well-defined, measure, $p$, that only depends on the data generated under one of the considered models, as opposed to $t$ which depends on both models and is usually set arbitrarily. 
\item Our solution is less sensitive to the curse of dimensionality as it does not require any of the $\Delta\{\eta(\bm{D}),\cdot \}$ to be close to 0. Indeed, our solution only considers the relative ranks of the $\Delta\{\eta(\bm{D}),\cdot \}$ calculated for the data generated under models 1 and 2.
\item Since our solution is less sensitive to the curse of dimensionality, it can accommodate vectors of summary statistics of any length. It only requires the design of a kernel function that ensures that the distributions of $\Delta\{\eta(\bm{D}),\cdot \}$ are well-separated under the competing models. This point is similar to the one made by \citet{Marin2014}, except that the separation, in our solution, can be studied on the real line as opposed to a high-dimensional space as suggested in \citet{Marin2014}.
\item Our solution uses the entire amount of pseudo-data generated and does so in a computationally efficient manner. 
\item Finally, our solution allows to formally preserve the user's priors on the model indices. 
\end{enumerate}

Our solution requires estimation of the ROC curve, which is the topic of the next two sections.
 
\subsection{Empirical ROC}
\label{section4.1}

We can leverage the relationship between Equations (\ref{eq:eq7}) and (\ref{eq:eq13}) to assign an ABC Bayes factor in several ways. Our first method is purely data driven and uses the following approximation of the ratio in Equation (\ref{eq:eq13}):
\begin{equation}
\label{eq:eq14}
	\dfrac{\widehat{\text{ROC}}(p)}{p}  = \dfrac{\dfrac{\sum^N_{i=1}\mathbb{I}[\mathcal{M}^{(i)}=1]\cdot\mathbb{I}[\Delta\{\eta(\bm{D}),\eta(\bm{D}^{\ast(i)})\}\leq t]}{N}}{\dfrac{\sum^N_{i=1}\mathbb{I}[\mathcal{M}^{(i)}=2]\cdot\mathbb{I}[\Delta\{\eta(\bm{D}),\eta(\bm{D}^{\ast(i)})\}\leq t]}{N}}.
\end{equation}
We see that by defining $t$ such that $\sum^N_{i=1}\mathbb{I}[\mathcal{M}^{(i)}=2]\cdot\mathbb{I}[\Delta\{\eta(\bm{D}),\eta(\bm{D}^{\ast(i)})\}\leq t]$ is constant for any $N$, and by increasing the number of simulations, the ratio in the denominator of Equation (\ref{eq:eq14}) will be driven to 0; hence, approximating the limit as $p \to 0^+$ in Equation (\ref{eq:eq13}). This approach has several major advantages as compared to current practice: 
\begin{enumerate}
	\item $t$ is chosen only as a function of the distance scores between the value of the summary statistic of the observed data and the value of the summary statistic of the data generated under model 2 (versus all distance scores in other implementations of the ABC algorithm).
	\item $t$ is chosen such that the number of accepted distance scores under model 2 is fixed (versus a fixed value of $t$ arbitrarily close to 0, or a varying value of $t$ based on a fixed quantile of the empirical distribution of all the scores from models 1 and 2 combined).
\end{enumerate}
For a given set of observed data, all current implementations of the ABC algorithm result in unpredictable variations of both the numerator and the denominator of the ratio in Equation (\ref{eq:eq7.5}) as $N$ increases, which makes its convergence difficult to monitor. By fixing the rate of convergence of the denominator in Equation (\ref{eq:eq7.5}), our approach has the potential to better plan computing resources and monitor convergence.

\subsection{The non-central dual beta ROC model}
\label{section4.2}

Our second approach extends further the relationship between the ABC Bayes factor and the ROC curve by noting that \citep{Pepe2003}:  
\begin{align}
	\label{eq:eq15}
	\lim_{p \to 0^+}\dfrac{\text{ROC}(p)}{p} &= \lim_{p \to 0^+} \frac{d}{dp}\text{ROC}(p) \\
	\label{eq:eq16}
	&= \lim_{p \to 0^+} \frac{d}{dp} F(G^{-1}(p)) \\
	\label{eq:eq17}
	&= \lim_{p \to 0^+} \dfrac{f(G^{-1}(p))}{g(G^{-1}(p))}
\end{align}
where $f(\cdot)$ and $g(\cdot)$ denote the probability density functions of distance scores under models 1 and 2, respectively. 

Assigning an ABC Bayes factor using Equation (\ref{eq:eq17}) requires evaluation of the first derivative of $\text{ROC}(p)$, as $p \to 0^+$, or the ratio of densities, $\frac{f(\cdot)}{g(\cdot)}$, as $p \to 0^+$. 

Following \citet{Metz1998}, \citet{MossmanPeng2016}, and \citet{ChenHu2016}, we choose to fit a semi-parametric model to the empirical ROC curve obtained from the finite number of $\Delta\{\eta(\bm{D}), \cdot\}$ generated by the algorithm. As noted by these authors, fitting a model to each of the score distributions makes the assumption that the scores follow this particular model. Instead, fitting a model directly to the ROC curve relies on the weaker assumption that there exists a monotonic transformation of the observed scores that results in scores whose distributions can be described by a simple model. Since the ROC curve is invariant to monotonic increasing transformations of the scores \citep{Pepe2003, Metz1998, MossmanPeng2016, ChenHu2016}, the simple model can be used to represent the ROC curve without knowledge of the underlying score distributions.

The common binormal representation of the ROC was considered \citep{Pepe2003}, however it can easily be shown that the limit of the binormal model as $p \to 0^+$ is either taking a value of 0, 1 or $\infty$. Instead we use a model based on two non-central beta distributions \citep{Johnson1995}. Placing a restriction on the first shape parameter of each distribution ($\alpha_F=\alpha_G=1$) results in the following semi-parametric model for the ROC curve
\begin{equation}
\label{eq:eq19}
	\text{ROC}(p)=F(G^{-1}(p \ | \ \alpha_G=1, \beta_G, \lambda_G) \ | \ \alpha_F=1, \beta_F, \lambda_F)
\end{equation}
where $F(\cdot)$ and $G(\cdot)$ are non-central beta distribution functions with parameters $\alpha_F$, $\beta_F$, $\lambda_F$, and $\alpha_G$, $\beta_G$, $\lambda_G$ corresponding to $F$ and $G$ respectively. Note that contrary to current ABC methods for model selection based on pattern recognition algorithm, our method only requires the estimation of four parameters. The restriction on the first shape parameter of the densities guarantees that their ratio, as $p \to 0^+$, produces a stable result for all parameter values within the support: 
\begin{equation}
\label{eq:eq20}
	\lim_{p \to 0^+} \frac{d}{dp}ROC(p)
	= \lim_{t \to 0^+} \frac{f(t)}{g(t)}
	= \dfrac{\exp{\{-\frac{1}{2}\lambda_F\}}}{B(1,\beta_F)} \dfrac{B(1,\beta_G)}{\exp{\{-\frac{1}{2}\lambda_G\}}}.
\end{equation}

Fitting the semi-parametric model requires the use of numerical optimisation techniques to estimate the parameter values. We use a two step fitting procedure:
\begin{enumerate}
	\item Obtain an initial set of parameter estimates for $\beta_F, \lambda_F,\beta_G, \lambda_G$ using a Maximum Likelihood Estimation approach for fitting a dual beta ROC model to continuous data \citep{Metz1998, MossmanPeng2016, ChenHu2016}.
	\item Refine these estimates using an objective function to minimise the $L^2$ norm between the semi-parametric model and the empirical ROC.	
\end{enumerate}

The first step of our procedure enables us to include information on which model produced each distance score $\Delta\{\eta(\bm{D}), \eta(\bm{D}^{\ast(i)})\}$. However, during our implementation we found that the fit of the semi-parametric model to the empirical ROC curve could be improved from this method. Meanwhile, the $L^2$ norm objective function alone did not allow us to fit an adequate model when the score distributions overlapped heavily. Combining both processes gave the best results. 

Once estimates for $\beta_F$, $\lambda_F$, $\beta_G$, and $\lambda_G$ are obtained, it is trivial to use Equation (\ref{eq:eq20}) to assign the ROC-ABC Bayes factor. In practice, when a limited number of distance scores near 0 are observed, the quality of the fit of the ROC model near 0 can produce Bayes factors of meaningless magnitude (e.g., larger than $10^{100}$ or smaller than $10^{-100}$); these Bayes factors would vary wildly from one computation to another using the same observed data. We found that bounding the rate of false positives, $p$, to some low value, such as $10^{-5}$, before evaluating Equation (\ref{eq:eq19}) produced more robust ABC Bayes factors.

\section{Weight of fingerprint evidence using ROC-ABC} 

ABC for model selection possesses some obvious similarities with the algorithm proposed by \citet{Neumann2012}. 
To assign a Bayes factor to an observed fingermark, Neumann et al.'s \citeyear{Neumann2012} algorithm considers two sets of scores. The first set contains scores measuring the level of dissimilarity between the observed fingermark and \textit{pseudo-fingermarks} generated by Mr. X. The second set contains scores measuring the dissimilarity between the observed fingermark and \textit{pseudo-fingermarks} originating from a sample of individuals from a population of potential alternative sources. The general idea behind \citet{Neumann2012}'s model is that, if $H_1$ is true, Mr. X. would generate many more pseudo-fingermarks that are similar to the observed fingermark than the individuals in the population of alternative sources. If we define:
\begin{enumerate}
	\item $\bm{D}$ using the observed fingermark, $\bm{e}_u$; 
	\item models 1 and 2 in Equations (\ref{eq:eq2}) and (\ref{eq:eq3}) as the prosecution model and the defence model, respectively;
	\item generative models corresponding to $M_1$ and $M_2$;
	\item a kernel function, $\Delta\{\cdot,\cdot\}$, to compare pairs of finger impressions;
\end{enumerate}
we obtain an ROC-ABC algorithm to approximate a Bayes factor for fingerprint evidence. The ROC-ABC algorithm for fingerprint evidence is summarised in Algorithm \ref{alg:ROCABCalg}. It will converge to the Bayes factor under the same conditions as discussed above (i.e. sufficient statistic across all models, optimal distance metric, infinite/large number of simulations, etc.). 

\begin{algorithm}
	\For{$i=1$ to $N$}{
		Sample a model index, $\mathcal{M}^{(i)}$, from $\pi(\mathcal{M}=m)$, where $m=1,2$\;
		\textbf{if $\mathcal{M}^{(i)}=1$:} Select the control print\;
		\textbf{else:} Sample a print from the alternative source dataset\;	
		Sample a set of distortion parameters\;
		Generate a pseudo-fingermark, $\bm{e}_u^{\ast(i)}$, by distorting the selected print using the sampled distortion parameters\;
		Compute $\Delta\{\eta(\bm{e}_u),\eta(\bm{e}_u^{\ast(i)})\}$\;
		}
	Assign the ROC-ABC Bayes factor using methods from Section \ref{section4.1} or \ref{section4.2}.
\caption{ROC-ABC algorithm for fingerprint evidence}
\label{alg:ROCABCalg}
\end{algorithm}

\subsection{Generation of pseudo-fingermark data}
\label{section5.1}

Implementation of the ROC-ABC algorithm for fingerprint evidence requires a model from which pseudo-fingermarks can be generated. We utilise the same fingerprint distortion model as \citet{Neumann2012} to generate pseudo-fingermarks from any given control print. This model mimics the way fingerprint features are displaced as the skin on the tip of a finger is distorted when pressed against a flat surface. The parameter space of the model represents a wide variety of distortion directions and pressures. The model allows many different distortions to be produced from a single finger. Our distortion model assumes that fingers distort in the same way, independently of factors related to the donor (e.g., age, weight, profession) and to the finger number (e.g., right vs. left hand, thumb vs. index finger). The model obviously does not cover all possible distortion and pressure conditions, and donor related factors; however, it is currently the only option to obtain a large number of pseudo-fingermarks from any given individual.  

\subsubsection{Generation of pseudo-fingermarks under the prosecution model}
\label{section5.1.1}

When comparing fingerprints, an examiner first detects $k$ features of interest on the fingermark. Second, the examiner compares it to the 10 control prints from the donor considered by the prosecution proposition and selects the finger appearing to be the most likely source of the mark. Finally, the examiner attempts to identify the most similar $k$ corresponding features out of the $n$ features present on the control print from the selected finger. Note that, once selected, the sets of features on the fingermark and the control print remain fixed for the duration of the experiment, and that the selection process results in a unique bijective pairing between the two sets of $k$ features. Our algorithm assumes that this selection and pairing process has been done before generating pseudo-fingermarks under $M_1$. When $M_1$ is selected by the algorithm, a pseudo-fingermark is generated from the $k$ features selected on the control print using the distortion model. By construction, the features on this pseudo-fingermark have the same pairing with the features of the fingermark as the ones on the control print. The uncertainty on the type of the feature was modelled as described in \citet{Neumann2015}. By repeating this process each time $M_1$ is selected, we can obtain a set of pseudo-fingermarks from the $k$ features selected on the control print of Mr. X.

\subsubsection{Generation of pseudo-fingermarks under the defence model}
\label{section5.1.2}

The defence proposition considers that, if the fingermark was not left by Mr. X, it must have been left by another person in a relevant population of alternative sources. Assuming that fingerprint patterns result from a completely random process during the development of the foetus, we can generate data under $M_2$, first, by randomly selecting an individual from any representative sample of donors from the human population, and secondly, by generating a pseudo-fingermark from this individual's $k$ minutiae configuration that is most similar to the $k$ minutiae observed on the fingermark. As in the previous section, the features of this  pseudo-fingermark have a unique bijective pairing with the features on the fingermark by construction. This process is repeated each time $M_2$ is selected to obtain a random set of pseudo-fingermarks from the population of potential donors considered by $M_2$. Since it would be unrealistic to repeat manually the selection of the most similar $k$ minutiae for each individual in a large sample, we use a commercially available automated fingerprint matching system to perform this task.

\subsection{Summary statistic}
Configurations of minutiae can be described numerically in the form of heterogeneous multi-dimensional random vectors containing the measurements summarised in Table \ref{table1}. 

\begin{table}
\caption{\label{table1} Variable types of various measurements from a friction ridge pattern.}
\centering
\fbox{
\begin{tabular}{l*{2}{c}}
	 & measurement & variable type\\
	\hline
	\rowcolor{white}
	minutia locations & Cartesian coordinates in pixels & continuous\\
	minutia orientations & angular measure & circular\\
	minutia types & bifurcation vs. ridge ending & discrete
\end{tabular}}
\end{table}

\begin{figure}[H]
	\centering	\makebox{\includegraphics[width=0.8\textwidth, bb=100 0 1500 500]{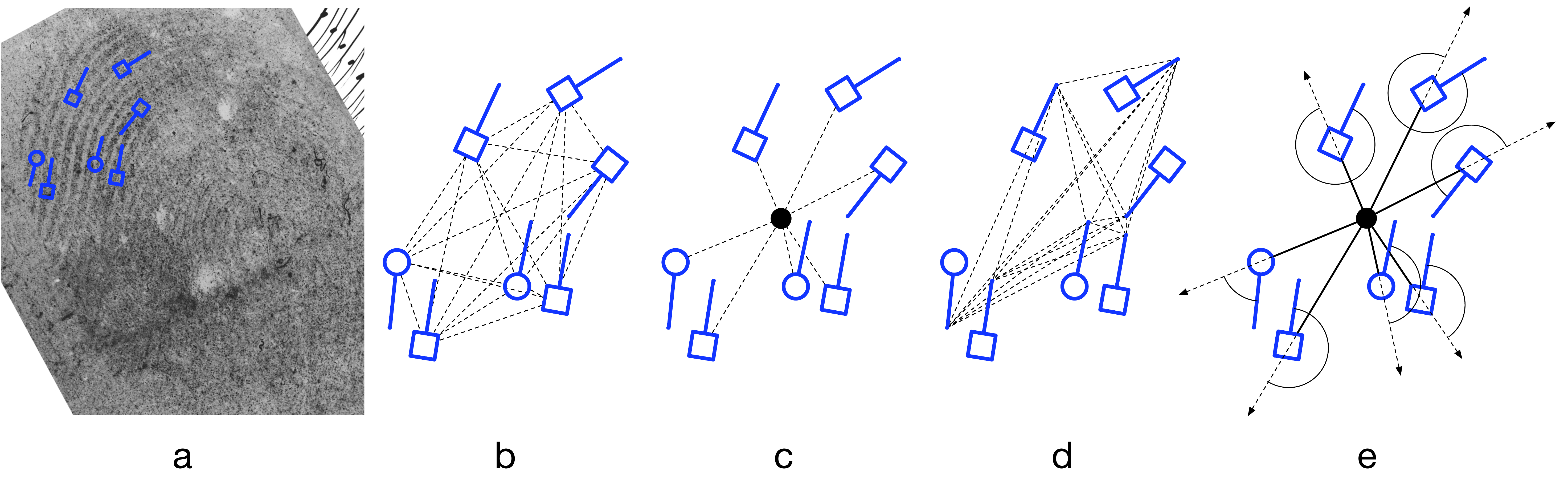}}
	\caption{\label{fig:fig1} (a) an annotated configuration of features on a friction ridge pattern. Squares and circles denote the types of the features and serve as markers for the locations of the features, while the extending line indicates the direction of the feature. (b) cross-distances between minutiae locations. (c) distances between minutiae locations and the centroid of the configuration. (d) cross-distances between the ends of the fixed-length segments originating from the centre of the minutiae and oriented in the same direction as the minutiae. (e) angles between the axes going from the centroid through the locations of the minutiae and the segments representing the minutiae directions are indicated by the solid line.}
\end{figure}

Minutiae locations and orientations are taken with respect to a coordinate system based on the frame of the impression's picture (Figure \ref{fig:fig1} (a)). Different framings of the same impression result in different measurements for the same set of features. For this reason, the original measurements need to be summarised in a way that is rotation and translation invariant. 
Several invariant measurements capturing the spatial relationships between the minutiae in a configuration can be calculated, such as the distances between every pair of minutiae in the configuration (Figure \ref{fig:fig1} (b)) and the distance of each minutiae from the centroid of the configuration (average of Cartesian coordinates of the minutiae in the configuration) (Figure \ref{fig:fig1} (c)). A similar approach can be used to define invariant summaries of the direction of each feature by using fixed-length segments to represent minutiae directions and by taking the distances between the ends of these segments for every pair of minutiae in the configuration (Figure \ref{fig:fig1} (d)). We also choose to represent minutiae directions as a function of the axes going from the centroid through the location of each minutia (angles are measured counterclockwise) (Figure \ref{fig:fig1} (e)). Feature types can be directly compared between configurations without the need to summarise since types are rotation and translation invariant.

All of these measurements can be brought together to create a vector of summary statistic of the original representation. For the example in Figure \ref{fig:fig1} with 7 features, a vector of summary statistic would include 21 cross-distances between pairs of minutiae, 7 distances between minutiae and the configuration's centroid, 21 cross-distances to capture the spatial representation of the minutiae directions, 7 angles and 7 types, for a total length of 63. For 10 features, the length of the vector of summary statistics would be 120; and for 15 features, it would be 255.

Given the heterogeneity and dimension of the measurements, it is unlikely that a sufficient summary statistic exists for fingerprint data. Here we adopt the approach which consists in pooling together as many individual summary statistics as possible in order to minimise the loss of information with respect to the original data. The curse of dimensionality is not a problem for our algorithm as discussed in Section~\ref{ROCmodifiedABC}. Furthermore, we note that, given that the model used to generate pseudo-fingermarks under $M_2$ is an extension of the model used under $M_1$, as our pooled summary statistic tends to sufficiency we can assume that it will be sufficient to compare between $M_1$ and $M_2$ \citep{Didelot2011}. Nevertheless, the vectors of summary statistics described above were designed to illustrate the concept of the ROC-ABC in the context of fingerprints and can certainly be improved upon through further investigation.

\subsection{Kernel function}
\label{kernel_function_Section}
Our ABC algorithm depends on a kernel function, $\Delta\{\cdot,\cdot\}$, which compares pairs of summarised configurations of minutiae. As mentioned in Section~\ref{ROCmodifiedABC} and according to~\cite{Marin2012}, we wish to use a kernel function that best separates the distributions of $\Delta\{\eta(\bm{e}_u),\cdot \}$ obtained under the competing models considered in Section~\ref{BF.fingerprint.evidence}. 

We developed a metric to compare pairs of summarised configurations of minutiae, and optimised the weights of several components to best separate the two distributions of $\Delta\{\eta(\bm{e}_u),\cdot \}$. For completeness, we have included this development and optimisation process in Appendix \ref{AppendixA}; however, we stress that other summary statistics, kernel functions and optimisation procedures could be considered without loss of generality of our proposed ROC-ABC method.

\subsection{Number of simulations}

For the purpose of this application, we limited the total number of pseudo-fingermarks generated for each test configuration to 500,000 (approximately 250,000 under each model). Assuming that we are interested in assigning the ROC-ABC Bayes factor for a specific fingermark in forensic casework, a much larger number of pseudo-fingermarks can be generated.

\section{Datasets}

The algorithm has been developed, optimised, and tested using several datasets. Each dataset comprises fingermarks or control prints captured digitally at 1:1 magnification and a resolution of 500 pixels per inch. All features were either labelled manually by an experienced fingerprint examiner or determined automatically by the encoding algorithm of the automated fingerprint matching system used in this study. The following features were extracted from every fingerprint used in the study:
\begin{enumerate}
\item the finger of origin of the impression, from 1 - right thumb - to 10 - left little finger (for control prints only);
\item the Cartesian coordinates of each minutia in pixels, using the bottom left of the image as the origin;
\item the direction of each minutia in radians, using the bottom left of the image as the origin and measuring counterclockwise; 
\item the type of each minutia: ridge ending, bifurcation or unknown;
\item the Cartesian coordinates of the centre of the impression (for control prints only). 	
\end{enumerate}
The four datasets (A, B, C, D) used in this study are described in Appendix \ref{AppendixB}.

\section{Results}

\begin{figure}
	\centering
	\includegraphics[width=0.7\textwidth, bb=200 0 700 400]{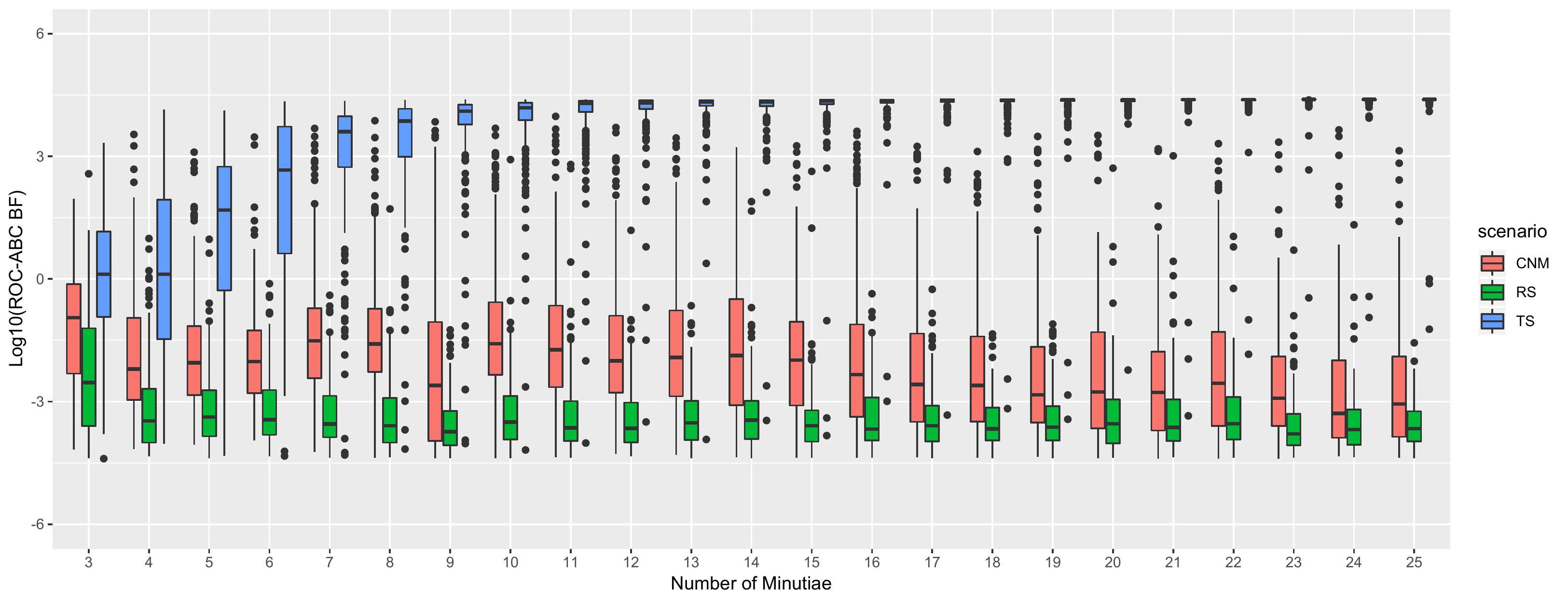}
	\caption{Results obtained using the empirical ROC method. Blue: results from the experiment where the control prints originate from the true sources (TS). Red: results when the control prints originate from sources with close non-matching prints (CNM). Green: results when the control prints originate from randomly selected sources (RS).}
	\label{fig:empirical_combined}
\end{figure}

\begin{figure}
	\centering
	\includegraphics[width=0.7\textwidth, bb=200 0 700 400]{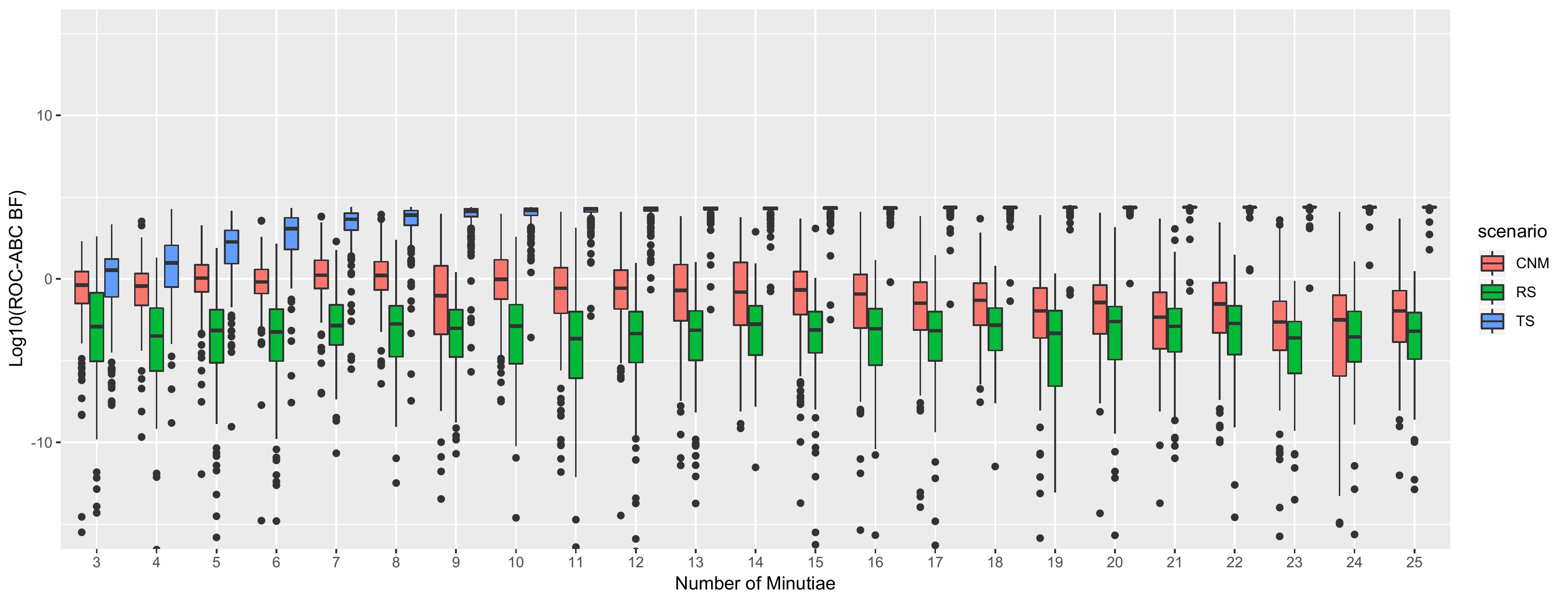}
	\caption{Results obtained using the non-central dual beta ROC model. Blue: results from the experiment where the control prints originate from the true sources (TS). Red: results when the control prints originate from sources with close non-matching prints (CNM). Green: results when the control prints originate from randomly selected sources (RS).}
	\label{fig:ROC_p_combined}
\end{figure}

Three experiments were performed using 4067 trace configurations ranging from 3 to 25 minutiae and sampled from 207 fingermarks. For each trace configuration, we consider in turn that:
\begin{itemize}[leftmargin=1.3cm]
\item[{\makebox[1.1cm]{TS:\hfill}}] Mr. X (the donor of the control configuration) is the true source of the trace configuration (dataset B);
\item[{\makebox[1.1cm]{CNM:\hfill}}] Mr. X is selected based on the similarity of his fingerprints to the trace configuration (dataset C);
\item[{\makebox[1.1cm]{RS:\hfill}}] Mr. X is randomly selected in a population of donors (dataset D).
\end{itemize}
In these experiments, the control configurations in datasets B to D were used to resample pseudo-fingermarks using $M_1$, and dataset A was used to resample pseudo-fingermarks using $M_2$. Results for all three experiments can be found in Figures \ref{fig:empirical_combined} and \ref{fig:ROC_p_combined}. ROC-ABC Bayes factors presented in Figure \ref{fig:empirical_combined} were assigned using the empirical ROC method discussed in Section \ref{section4.1}. Figure \ref{fig:ROC_p_combined} presents ROC-ABC Bayes factors that were assigned using the non-central dual beta ROC model described in Section \ref{section4.2}. Note that the vertical axis in \ref{fig:ROC_p_combined} has been truncated to focus on the mass of the distributions and that some extreme outliers may not be represented.

In experiment TS, the control prints originate from the true sources of the marks, and so $H_1$ is true.  Both figures show a similar behaviour where the magnitude of the ROC-ABC Bayes factor increases as the number of minutiae increases. In both cases, the ROC-ABC Bayes factors appear bounded. The bound for the empirical ROC-ABC Bayes factors stems from Equation (\ref{eq:eq14}), which requires us to set a value for $p$. In this case, we define $p$ as the false positive rate evaluated at the $10^{th}$ smallest $\Delta\{\eta(\bm{e}_u),\cdot \}$ generated under $H_2$. This corresponds to approximately 1 in 25,000. The bound for the non-central dual beta ROC model stems from using the same $p$ as in the empirical model. Bayes factors erroneously supporting $H_2$ are noted in both series of results for smaller configurations of minutiae (3 to 7 minutiae), which is not surprising as these configurations contain less discriminative information and are more likely to be observed in fingers from different individuals. Nevertheless, we note that only a handful of ROC-ABC Bayes factors support the wrong proposition for larger configurations. These cases deserve further investigation; at this time, we believe that they are related to configurations displaying unusual distortion that cannot be handled by the current generation of the distortion algorithm. Improvement of the summary statistic, kernel function, and distortion algorithm may be able to minimise further the number of cases where the Bayes factor erroneously supports $H_2$. Finally, we observe that the range of values taken by the ROC-ABC Bayes factor for different configuration sizes overlap which supports the observations made by \citet{Neumann2012} that there is no scientific justification for the use of a fixed number of minutiae as a decision point to distinguish between $H_1$ and $H_2$, and that the evidential value of each configuration needs to be quantified based on its own characteristics. 

In experiment CNM, the control prints originate from non-mated sources that are chosen due to their similarity to the traces, and so $H_2$ is true. In both Figures~\ref{fig:empirical_combined} and \ref{fig:ROC_p_combined}, we observe that a large proportion of the ROC-ABC Bayes factors erroneously support the hypothesis of common source, $H_1$, although the algorithm using the empirical ROC appears to perform significantly better. The high rate of misleading evidence is not a surprise for low numbers of minutiae since it is not difficult to find multiple similar configurations on different fingers in large a dataset. The high rate of misleading evidence is more surprising for larger configurations. Larger values of the ROC-ABC Bayes factor when $H_2$ is true occur when the kernel function used by the algorithm considers the pseudo-fingermarks generated using $M_1$ more similar to the observed fingermark than they really are. As mentioned before, improvement of the summary statistic and the kernel function should significantly reduce the rate of misleading evidence in favour of $H_1$. In practice, we believe that examiners comparing close non-matching finger impressions would be able to exclude that they originate from a common source by visual inspection using friction ridge characteristics that are not taken into account by our metric. 

In experiment RS, the control prints are obtained from randomly selected sources, and so $H_2$ is also true. In both cases the majority of observations correctly support the hypothesis of different sources, $H_2$. As in the second experiment, the algorithm using the empirical ROC curve significantly outperforms the one based on the non-central dual beta model of the ROC curve. 

Overall, we note that the semi-parametric modelling of the ROC curve needs to be improved. In many situations, it appears that the dual-beta model of the ROC curve is not optimal and that mixtures of beta distributions would better represent the data. 

These experiments were repeated using the logistic regression method by \citet{Beaumont2008} (see Appendix \ref{AppendixC}). Results can be found in Figure \ref{fig:logistic_combined} of Appendix \ref{AppendixC}. It is important to note that the logistic method does not present an obvious upper bound for the results in experiment TS and assigns Bayes factors with notably large magnitudes. This property is not desirable since it may lead to unrealistic magnitude of support. 

A comparison of the computational times for the empirical and parametric ROC methods and the logistic method is presented in Figure \ref{fig:time_combined} of Appendix \ref{AppendixC}. The empirical ROC-ABC method was without rival in terms of computation time. The logistic method performed at a much slower rate, and the difference between the two methods increases with the dimensionality of the data. When compared to a widely used ABC algorithm, we find that our method provides the high computational efficiency that is necessary to provide real-time calculation of the weight of forensic evidence in casework.

\section{Discussion}

The contribution of this paper is twofold. Firstly, we have proposed a method to rigorously quantify the weight of fingerprint evidence using the formal statistical framework provided by ABC. Secondly, our ROC-ABC algorithm helps with some of the issues commonly associated with ABC algorithms. 

Overall, our results are consistent with the results presented by \citet{Neumann2012} while capturing the user's belief about the parameters of the two competing models and providing an alternative to the weighting function that they proposed:
\begin{enumerate}
	\item The probability of misleading evidence in favour of the defence decreases dramatically as the number of minutiae increases.
	\item The probability of misleading evidence in favour of the prosecutor is very low for configurations greater than 7 minutiae, when the donor has been randomly selected. As expected, this probability is higher when the donor has been selected based on the similarity of its fingerprints with the fingermark. Improvements in the summary statistic used to describe fingerprint patterns and in the kernel function used to compare them, together with the ability of fingerprint examiners to account for more discriminative information than our model, will certainly reduce the rate of misleading evidence in favour of the prosecutor in an operational implementation of the model.
	\item The overlap between the ranges of values of the ROC-ABC Bayes factor across different numbers of minutiae confirms that the use of the number of corresponding minutiae is only one of the criteria for inferring the identity of the source, and that the contribution of additional information regarding fingerprint pattern needs to be taken into account when determining the donor of a fingermark. 
\end{enumerate}

\begin{figure}
	\centering	
	\includegraphics[width=0.8\textwidth]{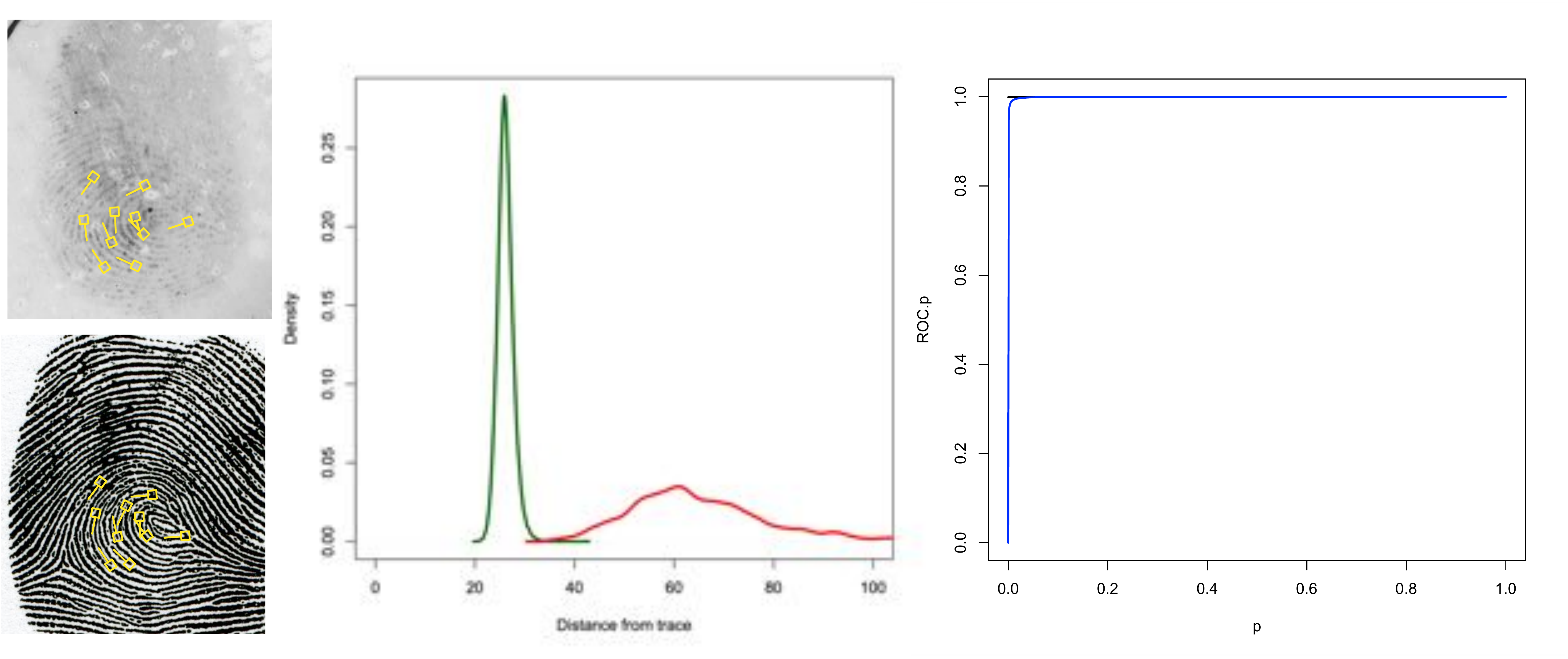}
	\caption{Left panel: Latent print (top) and control print (bottom), both originating from the same source. Corresponding features between the two impressions have been annotated in yellow. Middle panel: Kernel density estimates of the densities of the scores generated under $H_1$ (green) and $H_2$ (red). Note that the distribution of scores generated under $H_1$ is situated much closer to 0, but neither of the curves cover 0. Right panel: Empirical ROC curve (black) and parametric ROC curve (blue) generated from the distributions of scores in the middle panel. Note the steep slope of the curve near $p=0$.}
	\label{fig:TS_example}
\end{figure}

\begin{figure}
	\centering	
	\includegraphics[width=0.8\textwidth]{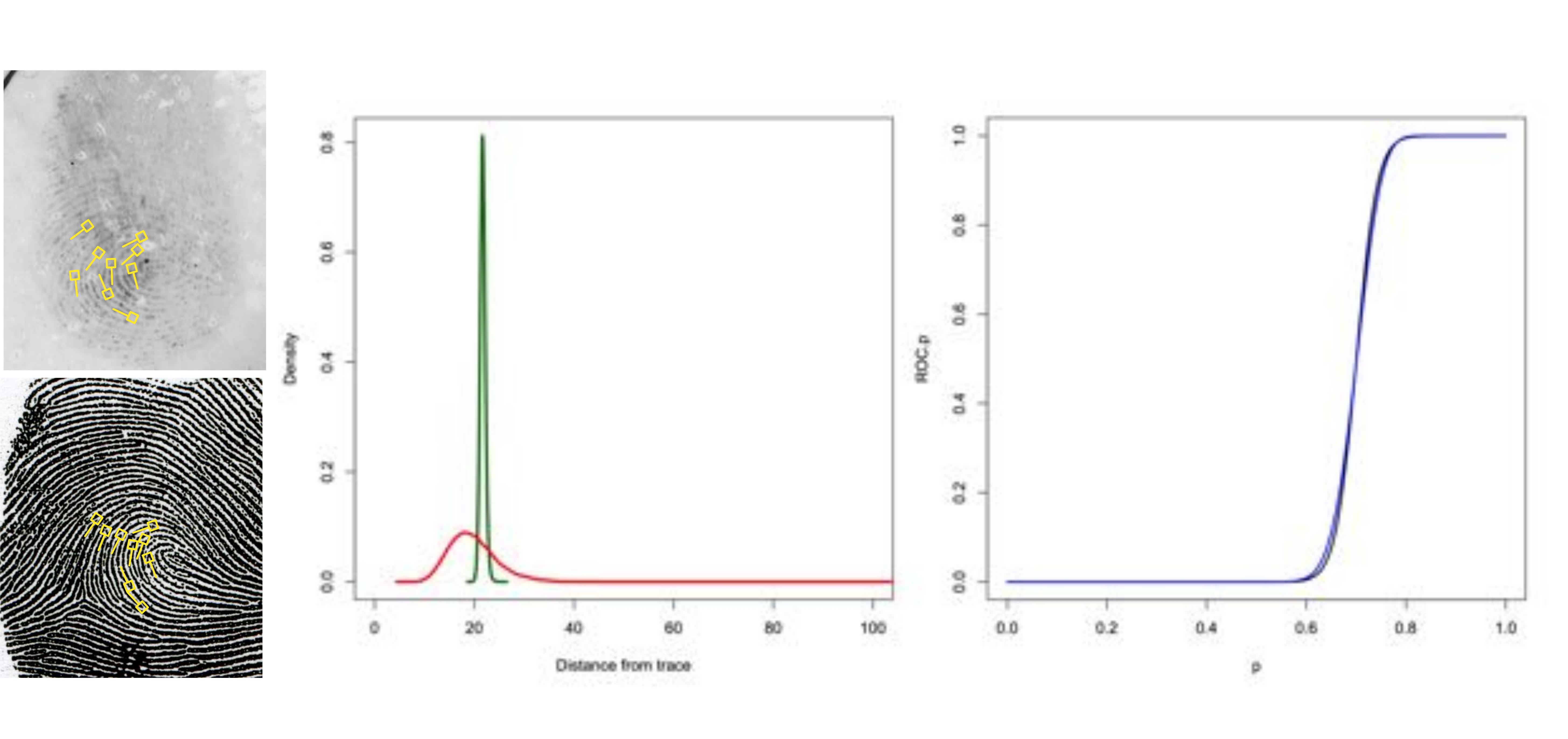}
	\caption{Left panel: Latent print (top) and control print (bottom) originating from the different sources. Potentially corresponding features between the two impressions have been annotated in yellow. Middle panel: Kernel density estimates of the densities of the scores generated under $H_1$ (green) and $H_2$ (red). Note that the algorithm has detected many more random impressions that are more similar to the latent print than the impression from the suspect. Right panel: Empirical ROC curve (black) and parametric ROC curve (blue) generated from the distributions of scores in the middle panel. Note the flat slope of the curve near $p=0$ indicating that the algorithm supports $H_2$.}
	\label{fig:DS_example}
\end{figure}

In addition to the straightforward operational implementation of our ROC-ABC approach, Figures \ref{fig:TS_example} and \ref{fig:DS_example} illustrate the powerful visual representation of the probative value of fingerprint comparisons that is offered by the ROC-ABC algorithm. In Figure \ref{fig:TS_example}, the latent print originates from same source as the control print and $BF_{abc}=2.51 \times 10^{7}$. The scores distributions and the ROC curve in the middle and right panels intuitively show that the data supports the hypothesis that the trace and control prints originate from the same source. In contrast, Figure \ref{fig:DS_example} shows a situation where the same latent print as in Figure \ref{fig:TS_example} is compared to a randomly selected control print. In this scenario, $BF_{abc}=3.56 \times 10^{-153}$. The clear support of the algorithm for the hypothesis that the latent and control impressions do not originate from the same source can be seen using the scores distributions and ROC curve in the middle and right panels. We believe that this intuition can easily be conveyed to jurors and other factfinders. 

We do not claim that the kernel function that was used in this paper is optimal. It is worth exploring adaptive kernels that maximise the separation between the pseudo-data generated by both models in any specific case. Nevertheless, while the summary statistic and the metric used to generate the results presented in this paper can be improved upon, they are adequate to show the potential of the concept of the ROC-ABC algorithm. Operational implementation of the method would require further studies of the repeatability of the values obtained by the algorithm as a function of different samples (and different sample sizes) of the population of potential donors considered by $H_2$. 

Our proposed modification to the standard ABC model selection algorithm results in several improvements. Our approach, based on properties of the ROC curve, transforms the convergence of the algorithm into a function of the rate of false positives in favour of the model considered in the numerator of the Bayes factor. Most current implementations of ABC for model selection result in unpredictable variations of both the numerator and the denominator of the ABC Bayes factor as the number of simulations, $N$, increases, which makes the convergence of the algorithm more difficult to monitor. In our approach, the tolerance level, $t$, for a given data set is chosen such that the number of accepted samples under the model considered in the denominator of the Bayes factor is fixed for all $N$. Hence, as $N$ increases, the approximation of the limit as the rate of false positives goes to 0 improves. Our approach has the potential to better plan computing resources. Critically, our method allows for rigorously monitoring convergence.

The shift from tolerance level on $\Delta\{\eta(\bm{D}),\cdot \}$ to rate of false positives in favour of model 1 does not require any of the $\Delta\{\eta(\bm{D}),\cdot \}$ to be close to 0. Instead, only the relative ranks of the $\Delta\{\eta(\bm{D}),\cdot \}$ calculated for the data generated under models 1 and 2 are considered. This implies that the kernel function used to assess level of similarity can accommodate vectors of summary statistics of any length, without encountering the curse of dimensionality because there is no need for any of the scores to be close to 0. Our algorithm only requires that the distributions of $\Delta\{\eta(\bm{D}),\cdot \}$ are well-separated under the competing models. This point is similar to the one made by \citet{Marin2014}, except that the separation, in our solution, can be studied on the real line as opposed to a high-dimensional space as suggested in \citet{Marin2014}. In addition to avoiding the curse of dimensionality, our method is also able to process the entire amount of pseudo-data generated in a computationally efficient manner and does not require filtering the pseudo-data: as the dimension of the summary statistic vector increases, the time required to assign Bayes factors using other methods (such as the logistic regression approach) increases exponentially (Figure \ref{fig:time_combined}). Finally, our solution allows to formally preserve the user's priors on the model indices. 

We have not formally addressed the issue of the sufficiency of the summary statistic that is required for the convergence of the ABC Bayes factor to the Bayes factor. This convergence is extremely important in some contexts, such as forensic science, where fact-finders are as equally interested in the proposition supported by the Bayes factor as in the magnitude of this support. Since the method that we propose is not affected by the curse of dimensionality, it permits including as much information in the summary statistic vectors as needed as in \citet{Pudlo2015}. 

To implement our approach in practice, we have proposed two methods to assign ROC-ABC Bayes factors. Our results show significant differences in performance between the empirical ROC and the dual beta ROC method. The empirical model appears to produce more stable and meaningful results (i.e., ABC Bayes factors with reasonable magnitude). As we increase the number of simulations, the empirical model naturally approximates the limit $p \to 0^+$ in Equation (\ref{eq:eq13}). The non-central dual beta ROC model has the potential to explore the limit when $p \to 0^+$ with a smaller sample size. In practice, we observed that the values obtained using Equation (\ref{eq:eq20}) for multiple runs of the ROC-ABC algorithm for a given set of observed data differ greatly from one another (many orders of magnitude on the $\log_{10}$ scale). It appears that Equation (\ref{eq:eq20}) is very sensitive to small changes in the values of the estimates of the model's parameters. Instead, we tried to fix $p=\frac{1}{25,000}$ in Equation (\ref{eq:eq19}) to obtain more robust values and generate the data in Figure \ref{fig:ROC_p_combined}. Our results show that in several cases our algorithm does not support the correct model; this may be due to our modelling of the ROC curve in the neighbourhood of 0 not being an accurate representation of the data. Once again, this shows that the non-central dual beta ROC model is very sensitive to small changes in the estimates of its parameters. Improvements can be made to the fitting procedure for the non-central dual beta ROC model, such as an explicit monotonic increasing transformation of the distance scores to initiate the numerical optimisation procedure with distributions that are closer to the assumed model; alternatively, other models whose limits at 0 exist can be investigated.

\section{Conclusion}

In this paper, we propose an algorithm to formally and rigorously assign Bayes factors to forensic fingerprint evidence. Our modified ABC model selection algorithm was used to address several criticisms of the model proposed by \citet{Neumann2012} by framing the problem into a formal Bayesian framework. The results presented here show that our method is promising, with low rates of misleading evidence, and has the potential to be applied to many other complex, high-dimension evidence forms such as shoe prints, questioned documents, firearms, and traces characterised by analytical chemistry. Ultimately, the widespread use of statistical approaches to quantify the weight of forensic evidence to replace the existing inference paradigm can only be enabled by technology providers offering commercial products to the forensic community. Our method leverages currently available technology that was designed to search forensic traces into large databases and retrieve the most likely candidates. For mainstream evidence types such as fingerprints, firearms, and shoe impressions, our algorithm can readily be implemented, validated, and integrated in current commercial offerings. Furthermore, we note that the use of ROC curves in the algorithm will be naturally familiar to engineers and scientists designing these systems, which may facilitate the implementation of our method in commercial systems. 



As an added benefit, our algorithm addresses several shortcomings of current ABC model selection methods. We use the properties of the Receiver Operating Characteristic curve to address the issue of choosing a suitable tolerance level when assigning ABC Bayes factors. Our modification allows for a natural convergence of the algorithm as the number of simulations increases, and for monitoring this convergence as a function of the sole rate of false positives in favour of the model considered in the numerator of the Bayes factor. Focusing on the rate of false positives (rather than  the tolerance level) allows our method to rely on the ordering of kernel scores, rather than the magnitudes of scores, and thus, is not subject to the curse of dimensionality. In addition, our method considers the entire amount of pseudo-data generated under the considered models in a computationally efficient manner.


\newpage

\appendix
\section{Kernel function development}
\label{AppendixA}

As mentioned in Section \ref{kernel_function_Section}, we wish to use a kernel function that best separates the distributions of $\Delta\{\eta(\bm{e}_u),\cdot \}$ obtained under the competing models considered in Section~\ref{BF.fingerprint.evidence}. Our kernel function, $\Delta\{\cdot,\cdot\}$, is a linear combination of several metrics, denoted by $\Delta_1\{\cdot,\cdot\}$, $\Delta_2\{\cdot,\cdot\}$, $\Delta_3\{\cdot,\cdot\}$, $\Delta_4\{\cdot,\cdot\}$, and $\Delta_5\{\cdot,\cdot\}$, corresponding to the different summary statistics described above and aimed at capturing differences in spatial relationships, directions, and types of the features: 
\begin{equation}
\label{eq:eq24}
	\Delta\{\cdot,\cdot\} = c_1\Delta_1\{\cdot,\cdot\} + c_2\Delta_2\{\cdot,\cdot\} + c_3\Delta_3\{\cdot,\cdot\} + c_4\Delta_4\{\cdot,\cdot\} + c_5\Delta_5\{\cdot,\cdot\},
\end{equation}
where $c_i$, for $i \in \{1,2,3,4,5\}$, are real-valued constants. The components of the metric are described below. We remind the reader that by construction (see Section \ref{section5.1}) the $i^{th}$ measurement in $\bm{e_u}$ is uniquely paired with the $i^{th}$ measurement in $\bm{e_u}^{\ast}$.

\subsection{Components of the distance metric}

The first component, $\Delta_1\{\cdot,\cdot\}$, captures the differences in cross-distances between the locations of the minutiae in a pair of configurations. Denoting the $i^{th}$ cross-distance from $\bm{e}_u$ by $d_i$, and the $i^{th}$ cross-distance from $\bm{e}_u^{\ast}$ by $d_i^{\ast}$, the first component of the distance metric is given by $\Delta_1\{\bm{e}_u,\bm{e}_u^{\ast}\} = \left( \sum_{i=1}^{\binom{k}{2}} \frac{(d_i-d^{\ast}_i)^2}{d_i} \right)^{1/2}$.
The second component, $\Delta_2\{\cdot,\cdot\}$, captures the difference in the spatial spread of the features. Denoting the distance of the $i^{th}$ minutiae from the centroid of the configuration from $\bm{e}_u$ by $d_i$, and the same from $\bm{e}_u^{\ast}$ by $d_i^{\ast}$, this component is given by $\Delta_2\{\bm{e}_u,\bm{e}_u^{\ast}\} = \left( \sum_{i=1}^{k} \frac{(d_i-d^{\ast}_i)^2}{d_i} \right)^{1/2}$.
The third component, $\Delta_3\{\cdot,\cdot\}$, takes the same form as $\Delta_1\{\cdot,\cdot\}$, but instead uses $d_i$ and $d_i^{\ast}$ as the $i^{th}$ cross-distance between location markers for feature directions (as illustrated in Figure \ref{fig:fig1} (d)) on $\bm{e}_u$ and $\bm{e}_u^*$ respectively. This component captures differences in directions of the features.
The fourth component, $\Delta_4\{\cdot,\cdot\}$, captures the difference in direction between the paired features for two configurations. Denoting the angle (measured in degrees) depicted in Figure \ref{fig:fig1} (e) for the $i^{th}$ minutiae of $\bm{e}_u$ by $\theta_i$, and the same from $\bm{e}_u^{\ast}$ by $\theta_i^{\ast}$, the fourth component is given by 
\begin{equation*}
\label{eq:eq27}
   \Delta_4\{\bm{e}_u,\bm{e}_u^{\ast}\}=
   \sum_{i=1}^k \left\{
	\begin{array}{ll}
	      \dfrac{\left|\theta_i - \theta_i^{\ast} \right|}{\theta_i} & \text{if } \left| \theta_i - \theta_i^{\ast} \right| \leq 180 \\
	      \dfrac{\left(180 - \left| \theta_i - \theta_i^{\ast} \right|\right) \mod 180}{\theta_i} & \text{if } \left| \theta_i - \theta_i^{\ast} \right| > 180 \\
	\end{array} 
	\right. .
\end{equation*} 

Finally, denoting the type of the $i^{th}$ minutiae from $\bm{e}_u$ by $\tau_i$, and the same from $\bm{e}_u^{\ast}$ by $\tau_i^{\ast}$, the fifth component is given by $\Delta_5\{\bm{e}_u,\bm{e}_u^{\ast}\} = \left(\sum_{i=1}^k \mathbb{I}[\tau_i = \tau_i^{\ast}] \right)^{1/2}$.

\subsection{Optimisation of the distance metric}
Since our algorithm can accommodate vectors of summary statistics of any length, we are not interested in performing any form of variable selection. However, we are interested in obtaining the ``best'' separation between the score distributions under the competing models in order to recover the correct model. This argument extend the one made by \cite{Marin2014}.

Values for all $c_i$ in Equation (\ref{eq:eq24}) can be obtained by maximising the separation between the distribution of $\Delta\{\cdot,\cdot\}$'s from minutiae configurations generated by the same donor, and the distribution of $\Delta\{\cdot, \cdot\}$'s from minutiae configurations generated by different donors. To obtain the results presented later in this paper, we used numerical optimisation to maximise the average area under 450 ROC curves obtained from configurations with 5, 8, 12, 17, and 23 minutiae. Each ROC curve was based on 50,000 distance scores as calculated in Equation \ref{eq:eq24} and obtained by comparing $k$ minutiae on a fingermark to pseudo-fingermarks resulting from the distortion of the true source of the fingermark (Section \ref{section5.1.1}) and from the distortion of other fingers (Section \ref{section5.1.2}). Our results indicated that the second and fifth components of our metric did not help discriminate between same-source and different-sources distance scores, and were assigned $c_2=c_5=0$. Our results also indicated that component 3 was the most useful to maximise the average area under the ROC curve ($c_3 = 6.5$), followed by component 1 ($c_1 = 1$) and by component 4 ($c_4 = 0.1$). 

We stress that other summary statistics, kernel functions and optimisation procedures could be considered without loss of generality of our proposed ROC-ABC method.

\section{Datasets}
\label{AppendixB}

The data used in this study includes four datasets, described below.

\subsection{Dataset A: relevant population of potential donors}

A dataset of control prints taken under controlled conditions from more than 400,000 individuals (only identified through randomly assigned ID numbers) was used as a sample of a population of potential sources. The size of the dataset was not driven by scientific considerations, but corresponds to the number of control prints that the authors managed to gather for research purposes. Locations, directions, and types of fingerprint features were extracted automatically. This dataset was used to generate pseudo-fingermarks under the defence model (see Section \ref{section5.1.2}) and to generate special test cases under the prosecution model (see below).

\subsection{Dataset B: True source dataset}

A dataset of 207 pairs of fingermark and control print were obtained from casework archives. While the sources of the fingermarks are unknown, a trained fingerprint examiner deemed that, within each pair, the fingermark and control print originate from the same finger. Since the fingermarks originate from casework, they were developed on multiple surfaces using different physicochemical methods and represent a range of fingermarks that can be observed in casework. The fingerprint features on each pair of impressions were manually annotated by the fingerprint examiner. Corresponding features between paired impressions were manually designated by the fingerprint examiner. 
 
Test configurations, ranging from 3 to 25 minutiae, were sampled from the fingermarks (one configuration of each size per fingermark). For each test configuration, the corresponding minutiae on the control print were also selected, thus providing pairs of ``matching'' configurations truly originating from the same source (or assumed to be). 

The fingermarks described here were used as $\bm{e}_u$ when generating the results presented below. The corresponding control prints were used as control prints from Mr. X to test the ROC-ABC algorithm when the prosecution proposition, $H_1$, is true.
\begin{table}
\label{table2} 
\caption{Number of test configurations for each number of minutiae in dataset B.}
\centering
\fbox{%
\begin{tabular}{l|*{10}{c}}
\rowcolor{Grey}
	\# of minutiae & 3 & 4 & 5 & 6 & 7 & 8 & 9 & 10 & 11 & 12 \\
\rowcolor{white}
	\# of configurations & 207 & 207 & 207 & 205 & 203 & 203 & 199 & 197 & 190 & 187 \\
\rowcolor{Grey}
	\# of minutiae & 13 & 14 & 15 & 16 & 17 & 18 & 19 & 20 & 21 & 22 \\
\rowcolor{white}
	\# of configurations & 184 & 179 & 174  & 173 & 170 & 168 & 162 & 154 & 151 & 144 \\
\rowcolor{Grey}
	\# of minutiae & 23 & 24 & 25 & & & & & & & total\\
\rowcolor{white}
	\# of configurations & 142 & 133 & 128 & & & & & & & 4067\\
\end{tabular}}
\end{table}

\subsection{Dataset C: ``Close non-matching" source dataset}

The 4067 test configurations generated from the 207 fingermarks from dataset B were automatically compared against the control prints from all individuals in the sample from the population in dataset A. The matching algorithm was set up to select the control configuration that appeared most similar to each test configuration.

These pairs of configurations were used to test the ROC-ABC algorithm when the defence proposition, $H_2$, is true and Mr. X is not the true source of the fingermark, but possesses a very similar configuration of minutiae. This is equivalent to testing the algorithm under the ``worst possible case" scenario, which may correspond to the situation where Mr. X has been brought to the attention of the prosecution as a result of a database search. 

\subsection{Dataset D: ``Random" source dataset}

The experiment performed to construct dataset C was repeated. However, in this case, the matching algorithm was set up to return random configurations from any of the control prints, regardless of their levels of similarity to the test configurations. 
 
These pairs of configurations were used to test the ROC-ABC algorithm when the defence proposition, $H_2$, is true and Mr. X is not the true source of the fingermark, and he has not been selected based on the characteristics of his fingerprints. This is equivalent to testing the algorithm in situations where Mr. X has been brought to the attention of the prosecution as the result of evidence unrelated to fingerprint.

\section{Results using \cite{Beaumont2008}'s method}
\label{AppendixC}

Results for all three experiments were also generated using the logistic regression method proposed by \cite{Beaumont2008} since this method is the basis for most ABC model selection algorithms and has been found to be working well. We use the logistic regression method with the same metric as the one used in our ROC-based ABC algorithm in order to have directly comparable results. We want to re-emphasise that it is critical for the forensic application to have a fair idea of the magnitude of support for a given model, and not only to be able to select the correct one. Thus, we did not compare our method to machine-learning-based ABC methods since they seem to be only focusing on the posterior distributions for the models \cite{Pudlo2015}, while Equation \ref{eq:eq13} shows that our method can converge to the Bayes factor of interest. 

\begin{figure}
	\centering
	\includegraphics[width=0.7\textwidth, bb=200 0 700 300]{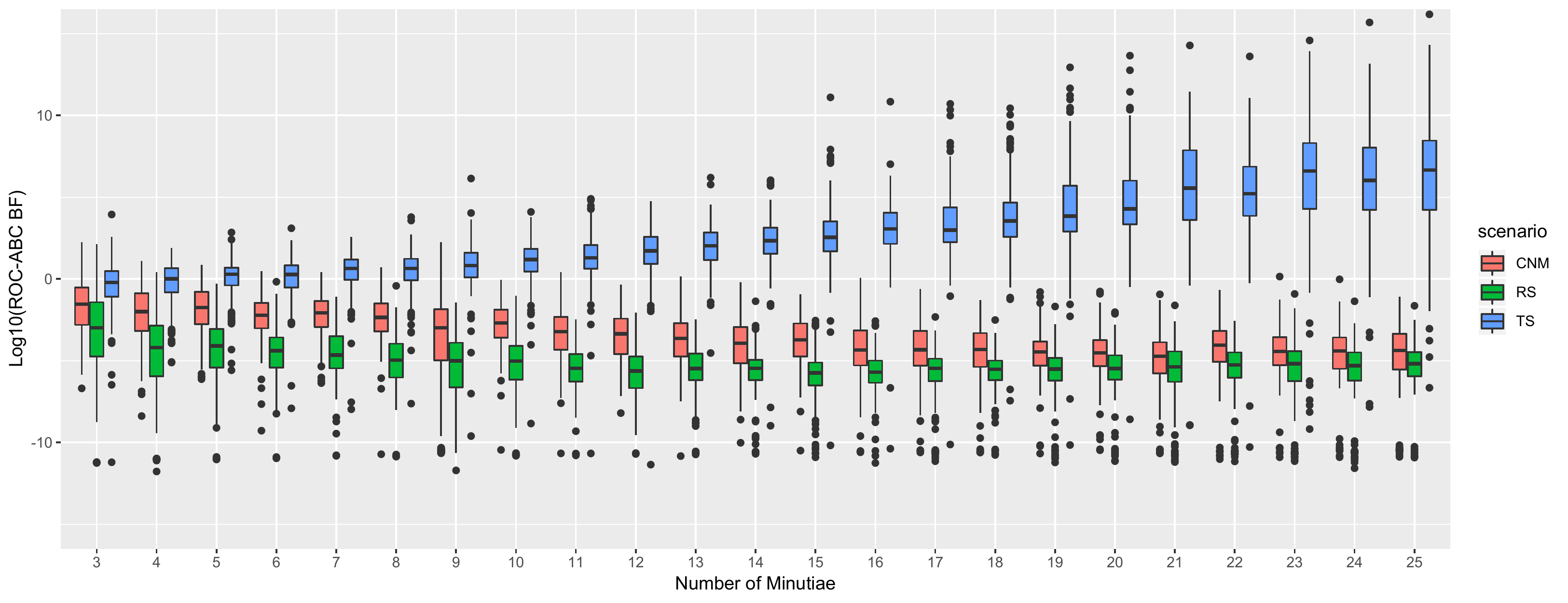}
	\caption{Results obtained using the logistic regression method. Blue: results from the experiment where the control prints originate from the true sources (TS). Red: results when the control prints originate from sources with close non-matching prints (CNM). Green: results when the control prints originate from randomly selected sources (RS).}
	\label{fig:logistic_combined}
\end{figure}

\begin{figure}
	\centering
	\includegraphics[width=0.7\textwidth, bb=200 0 700 300]{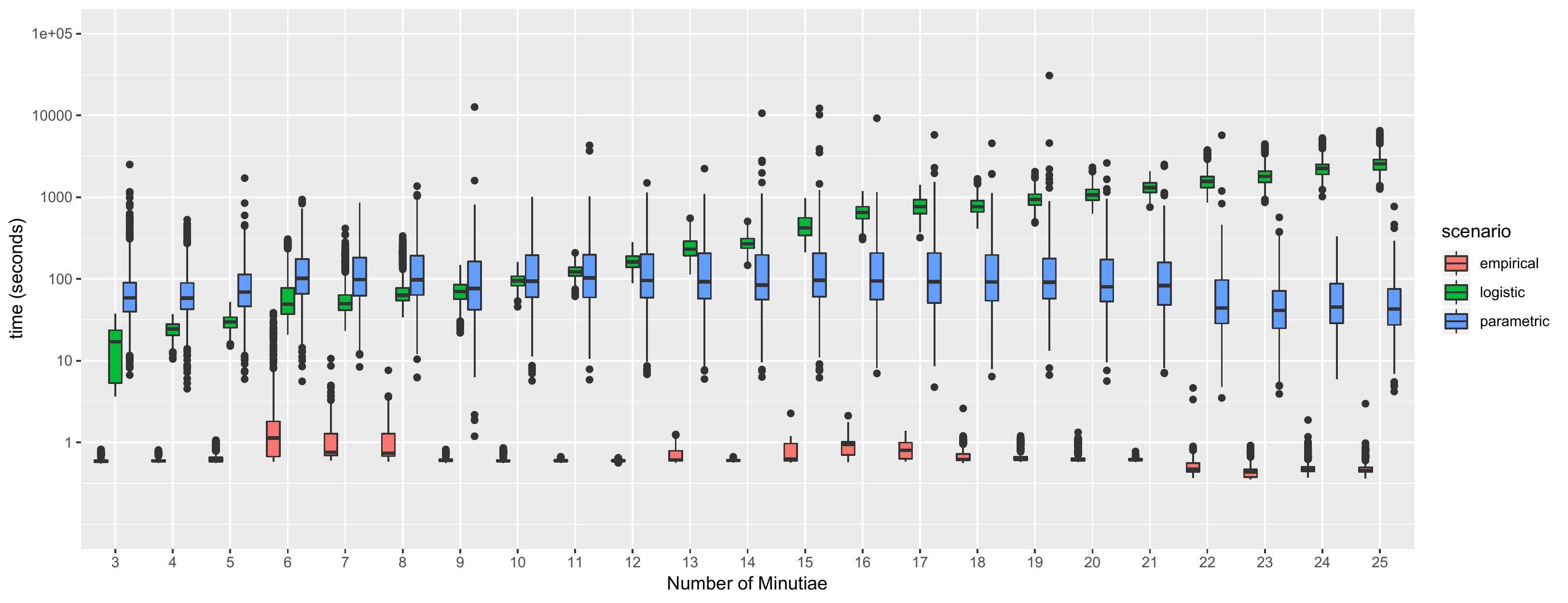}
	\caption{A comparison of the computation time for the empirical ROC method (red), the parametric ROC model (blue), and logistic regression method (green). The computation time represents the time required to assign Bayes factors once the pseudo-data has been generated.}
	\label{fig:time_combined}
\end{figure}

These results are presented in Figure \ref{fig:logistic_combined}. A comparison of the computation times for each of the three methods (empirical ROC, parametric ROC, and logistic regression) is presented in Figure \ref{fig:time_combined}.

The general trend of the results of the logistic regression method are similar to those of the ROC-based methods. For the experiment TS, the magnitude of the ABC Bayes factor increases as the number of minutiae increases, while for experiments CNM and RS, the ABC Bayes factors tend to generally support $H_2$.

However, the logistic method does not present an obvious upper bound for the results in experiment TS and assigns Bayes factors with notably large magnitudes. Based on the convergence results shown in Equations~(\ref{eq:eq7}) to (\ref{eq:eq13}), we do not believe that the larger magnitude of the Bayes factors in Figure~\ref{fig:logistic_combined} can be justified by the number of simulations performed in this experiment. We can only conclude that these Bayes factors severely overstate the weight of the evidence observed and generated in these cases. In addition, we note the very large variance of the Bayes factor assigned using the logistic regression method during experiment TS. This large variance results in a high rate of misleading evidence in favour of $H_2$. This rate is noticeably greater than that from the empirical method. 

Interestingly, results from experiments CNM and RS show that the logistic regression method produces less misleading evidence in favour of $H_1$ when $H_2$ is true, even when very similar prints are used. The logistic regression method maximises the separation between the two models by leveraging all of the content of the vectors of summary statistics, while the kernel function described in Section~\ref{kernel_function_Section} has been optimised for the average case. The ROC-ABC method may be improved in this aspect by using an adaptable kernel function that would also maximise the separation in each case. 

Overall, while using the logistic regression method as proposed by \cite{Beaumont2008}, we noted that the weighting of the pseudo-data prior to fitting the logistic regression model resulted in the removal of large portions, if not all, of the data generated under either $M_1$ or $M_2$. As discussed previously, this results in altering the user-defined priors on the model index and replacing them by unpredictable data-driven priors. Furthermore, this led to instability when fitting the logistic regression model.

A comparison of the computation time among the three methods (empirical ROC, parametric ROC, and logistic regression) is presented in Figure \ref{fig:time_combined}. This computation time represents the time require to assign ABC Bayes factors using the three different methods once the pseudo-data has been generated. The empirical ROC-ABC method was without rival in terms of computation time. Even as data complexity/dimensionality increased, computation time was relatively constant. The logistic method outperformed the parametric ROC-ABC method up until 9 minutiae. At this point, the total computation time for the logistic method continued to increase at an exponential rate while the computation time for the parametric ROC-ABC method remained fairly uniform. This is unsurprising since the computational complexity of the parametric ROC-ABC is driven by the number of univariate scores in the ROC curve, and not by the dimension of the vectors of summary statistics. In addition to an increase in computing time, an increase in computing resources was also required by the logistic regression method such that fingerprint data with more than 22 features could not be processed on a standard desktop computer, while the ROC-based methods have a very small computing footprint (once the initial pseudo-data has been generated).

\end{document}